\documentclass[aps,prl,superscriptaddress,twocolumn,amsmath,amssymb]{revtex4-2}
\pdfpageattr {/Group << /S /Transparency /I true /CS /DeviceRGB>>}

\usepackage{graphicx}
\usepackage{color}
\usepackage[hidelinks]{hyperref}
\usepackage{orcidlink}
\usepackage{amsmath}
\usepackage{amsthm}
\usepackage{physics}
\usepackage{amsfonts}
\usepackage{times,txfonts}
\usepackage[normalem]{ulem}

\hypersetup{colorlinks=true, linkcolor=black, citecolor=black, urlcolor=blue}

\begin{document}
\author{Christoph Hotter\,\orcidlink{0009-0003-3854-0264}}
\email[]{christoph.hotter@nbi.ku.dk}
\affiliation{Institut f\"ur Theoretische Physik, Universit\"at Innsbruck, Technikerstra{\ss}e\,21a, A-6020 Innsbruck, Austria}
\affiliation{Niels Bohr Institute, University of Copenhagen, Blegdamsvej 17, Copenhagen DK-2100, Denmark} 
\author{Arkadiusz Kosior\,\orcidlink{0000-0002-5039-1789}}
\affiliation{Institut f\"ur Theoretische Physik, Universit\"at Innsbruck, Technikerstra{\ss}e\,21a, A-6020 Innsbruck, Austria}
\author{Helmut Ritsch\,\orcidlink{0000-0001-7013-5208}}
\affiliation{Institut f\"ur Theoretische Physik, Universit\"at Innsbruck, Technikerstra{\ss}e\,21a, A-6020 Innsbruck, Austria}
\author{Karol Gietka\,\orcidlink{0000-0001-7700-3208}}
\email[]{karol.gietka@uibk.ac.at}
\affiliation{Institut f\"ur Theoretische Physik, Universit\"at Innsbruck, Technikerstra{\ss}e\,21a, A-6020 Innsbruck, Austria}

\title{Conditional  {Entanglement Amplification} via  {Non-Hermitian} Superradiant  {Dynamics}}

\begin{abstract}
Due to the inherently probabilistic nature of quantum mechanics, each experimental realization of a dynamical quantum system  {can produce distinct measurement outcomes, particularly when coupled to a dissipative environment.} Although  {quantum trajectories that lead to exotic, highly entangled states are possible in principle, their observation is typically hindered by extremely low probabilities.} 
In this work, we present a method to  {significantly enhance the probability of generating highly entangled states} in an ensemble of atoms undergoing  {collective superradiant decay on timescales much shorter than the individual atomic spontaneous emission rate.} 
By analyzing an effective non-Hermitian Hamiltonian governing the dynamics between photon emission events, we identify the conditions necessary for  {these rare no-click trajectories to occur with higher likelihood. Crucially, our method relies on initializing the system in a non-classical state, whose entanglement is amplified via the non-Hermitian superradiant dynamics.} 
This approach provides a  {new route to creating highly entangled macroscopic states such as atomic Schrödinger cat states, paving the way for advancements in quantum metrology and other quantum technologies.}
\end{abstract}
\date{\today}
\maketitle




 {\emph{Introduction---}}{Highly entangled macroscopic states play a key role in quantum-enhanced technologies such as quantum simulation and computing~\cite{Feynman1982,Ladd2010,Hauke2012,Georgescu2014}, quantum teleportation~\cite{Bennett1993,Pirandola2015}, high-precision spectroscopy~\cite{leibried2004highprecisoinspectro, colombo2022timereversal, Li2023}, and error-correcting codes~\cite{Hamming1950,moon2020error}. As a result, there has been significant experimental and theoretical effort to develop methods for their generation~\cite{grangier2007numberstatecat,Merkel_2010,liao2016catstate,serikawa2018catstateoptical,Anna2018Noonstate,chen2019heraldedNOON,takase2021noonsbustraction,weiQin2021longlivedcat,entanglementamplification2021jizohong}. A promising approaches to create these highly entangled states is one-axis twisting (OAT)~\cite{ueda1993SqueezedSpinStates,Wineland1994}, which can be realized in various physical platforms~\cite{MA201189squeezing,liu2018squeezincavityqed,Plodzien2020,gietka2021oatdistinctH,Plodzien2022,tana2023squeezingOpenchains,rydberg2023squeezingYAO,rydbergsqueeze2023JunYe,Dziurawiec2023,rydbergcats2024JunYe,Plodzien2024}. These platforms often rely on interaction-mediating excitations, such as photons in cavity-QED experiments~\cite{molmer2002badcavities,Borregaard_2017,braverman2019unitary,pedrozo2020entanglement,colombo2022timereversal} or phonons in ion systems~\cite{molmersorensongat1999}.}
\begin{figure}[htb!]
    \centering
    \includegraphics[width=0.49\textwidth]{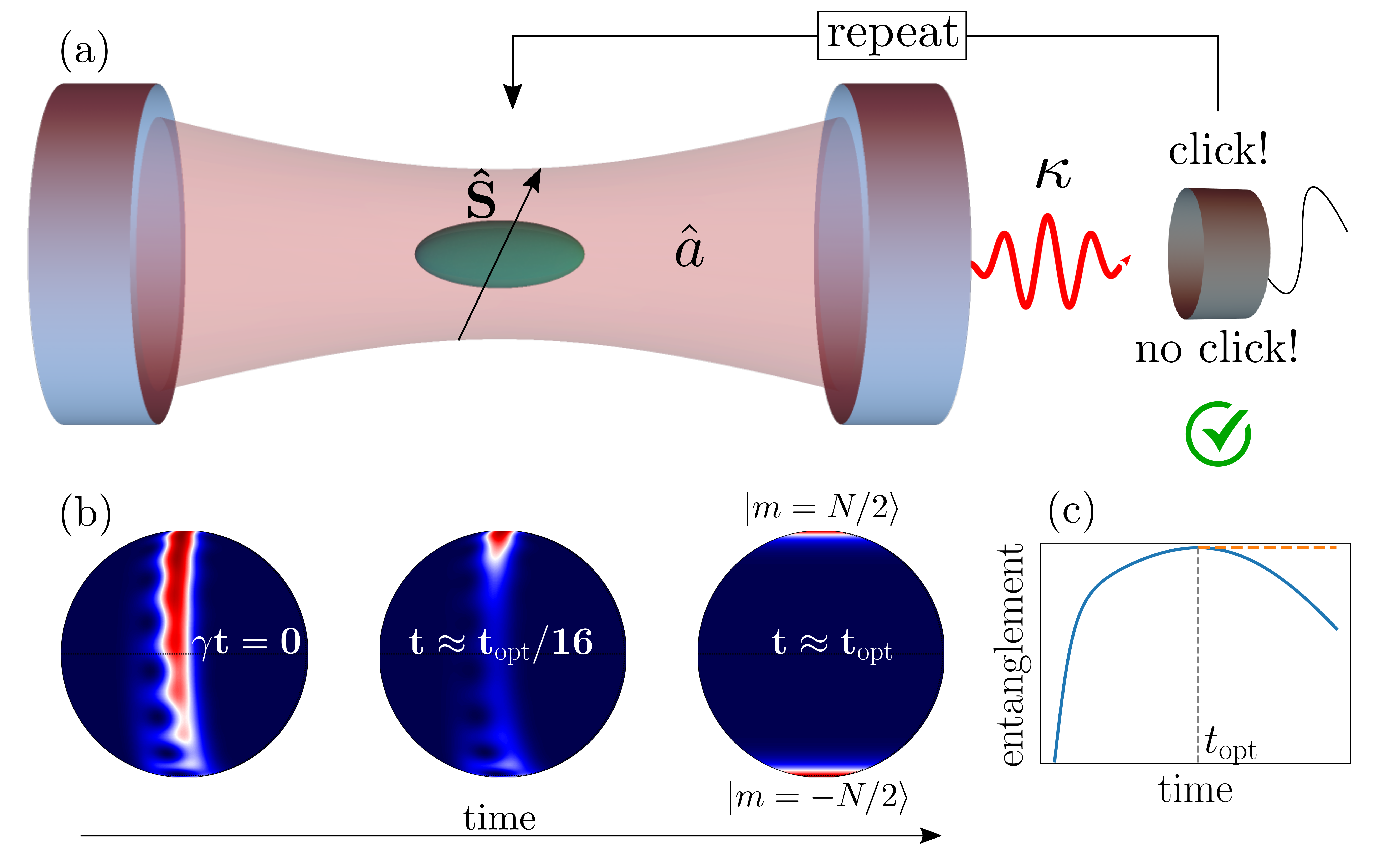}
    \caption{ {Schematics of the experimental setup and post-selection protocol. (a) The considered system consists of an ensemble of $N$ identical two-level atoms described with a collective spin operator $\hat {\mathbf S}$ interacting with a single-mode ($\hat a$) optical cavity with damping rate $\kappa$, described by the Tavis-Cummings model, see Eq.~\eqref{eq:TC_model}. (b) In the no-click limit, where no photon is being emitted, quantum trajectories of an atomic ensemble coupled to a cavity mode can evolve through exotic quantum states as shown in the main text. By preparing a suitable initial state, the probability of these trajectories can be greatly increased leading to the conditional generation of maximally entangled cat states (most right Bloch sphere). The no-click limit, can be achieved by using post-selection: if a photon is being observed during the time evolution, the experimental run needs to be repeated. (c) After a chosen evolution time--which can be optimized to reach the highest entanglement ($t_\mathrm{opt}$)---the atoms are decoupled from the cavity to stop the superradiant dynamics.
    } 
    } 
    \label{fig:setup_sketch}
\end{figure} 
Of particular interest are maximally entangled states, such as the Schr\"odinger cat state, 
which is a quantum state composed of a superposition of macroscopically distinguishable configurations. 
In a more down-to-earth physical example, a cat state can be represented by a superposition of \(N \gg 1 \) two-level atoms all occupying the same single-particle state~\cite{Singh1997atomiccats} (either \( |\!\downarrow\rangle \) or \( |\!\uparrow\rangle \)) as  
%
\(
    |\mathrm{cat}\rangle = \frac{1}{\sqrt{2}} \left(|\!\uparrow\rangle^{\otimes N} + e^{i \phi}|\!\downarrow \rangle^{\otimes N}\right),
\)
with \(\phi\) being the relative phase between the cat's components. This state represents a generalized Greenberger--Horne--Zeilinger state~\cite{Greenberger2009}, closely related to the maximally entangled NOON state~\cite{silberberg2019noonlight}. 
 {However, cat states are notoriously fragile, making them highly susceptible to decoherence. These features have so far limited the creation of truly macroscopic cat states.} 
Current state-of-the-art experiments have produced cat states involving around 20 qubits~\cite{Song2019catstate20qubits,lukin2019cat,blatt2021cat24}, with the largest achieved state comprising 32 ions~\cite{Moses2023}.  {These limitations highlight the urgent need for robust and alternative methods to generate maximally or highly entangled states, particularly approaches that harness dissipation in innovative ways}~\cite{rajagopal2001deco,braun2002entcommonbath,knight2002entheatenv,milburn2002steadystateent,hormeyll2009environentan,nemoteo2018negativetempent,dias2023entanglemetndisipation}. 



Dissipative quantum systems are most commonly described using the Lindblad master equation. Unlike the Schrödinger equation, which deals with the evolution of a single wavefunction, the Lindblad master equation describes the time evolution of a density matrix, $\hat \rho$, which, in general, represents a statistical ensemble of wavefunctions.  {An alternative approach to modeling the dynamics of open quantum systems is the Monte-Carlo wavefunction method~\cite{dalibard1992wavefunction, dum1992monte, molmer1993monte}.} In this method, at each time step, a quantum jump (i.e., a sudden, discontinuous change in the quantum state) may occur with a probability determined by the state of the trajectory.  {Between jumps, the wavefunction evolves according to a Schrödinger equation governed by a non-Hermitian Hamiltonian,}
\(
    \hat{H}_\mathrm{nh} = \hat{H} - \frac{i \hbar}{2} \sum_i \hat{J}_i \hat{J}_i^\dagger,
\)
where $\hat{H}$ is the usual Hermitian part and $\hat{J}_i$ are the jump operators. The jump probability is determined by the norm of the wavefunction, which decreases over time due to non-Hermitian evolution. Since the stochastic average over sufficiently many Monte-Carlo trajectories converges to the full solution of the Lindblad master equation, both methods yield identical averages for measurement outcomes.  {Although often viewed as a mathematical tool for handling large systems, the Monte-Carlo wavefunction method provides an intuitive physical interpretation:} trajectories can represent actual realizations of ideal measurements performed during individual experimental runs~\cite{breuer2002theory,Daley04032014,Lev2024replicaQT}.  {This approach inherently accounts for the non-deterministic nature of quantum jumps.} That being said, a single run (either experimental or numerical) can deviate significantly from the most probable outcome dictated by the underlying probabilities. In particular, some runs may lead to highly entangled quantum states. These trajectories, however, are typically rare, leaving their  {systematic} exploitation largely unexplored.

In contrast, in this work, we not only demonstrate that the efficient identification of rare quantum trajectories is possible via post-selection, but  {we also show how to significantly enhance their likelihood by optimizing the initial state.} Specifically,  {we illustrate this using} a non-Hermitian description of cavity superradiance~\cite{dicke1954coherence,norcia2016superradiance,norcia2018frequency,laske2019pulse,hotter2023cavity,bohr2024collectively},  {demonstrating} how the collective coupling of an atomic ensemble to a single mode of a lossy cavity can lead to the conditional generation of highly entangled states, including maximally entangled atomic cat states  {with a finite probability on timescales much faster than the spontaneous emission rate.}


 {\emph{Entanglement amplification via superradiance---}}The interaction between an ensemble of identical two-level atoms with a single-mode optical resonator (see Fig.~\ref{fig:setup_sketch}a) is described by the Tavis-Cummings model~\cite{helmut2013rmp}. In the reference frame rotating with the frequency of the cavity mode $\omega_c$, the Hamiltonian reads
\begin{align}\label{eq:TC_model}
    \hat H = -\Delta \, \hat S_z + g \left(\hat a \hat S_+ + \hat a^\dagger \hat S_-\right),
\end{align}
where $\Delta  \, \equiv \omega_c - \omega_a$ is the atom-cavity detuning, $g$ is the effective atom-cavity coupling, $\hat S_z$, $\hat S_+$, $\hat S_-$ are collective spin operators describing the ensemble of $N$ two-level atoms, and $\hat a$ ($\hat a^\dagger$) is the annihilation (creation) operator of a cavity photon. In the regime where the cavity damping rate $\kappa$ is much larger than the effective collective coupling rate $\sqrt{N} g$ (superradiant regime), the cavity mode can be adiabatically eliminated. For $\Delta = 0$, this leads to an effective master equation for the atomic degree of freedom~\cite{orioli2022emergent} which is solely described by collective decay 
\begin{align}
     \dot {\hat \rho} = - \gamma\left(\hat S_-\hat \rho \hat S_+ -\frac{1}{2}\left\{\hat \rho, \hat S_+\hat S_-\right\} \right),
     \label{eq:coll_decay}
\end{align}
 where $\gamma = g^2/\kappa $ is the effective collective atomic emission rate. Note that for a non-vanishing detuning, an additional dipole-dipole interaction term appears in the effective Hamiltonian~\cite{orioli2022emergent}.  {The following analysis merely uses the collective decay described in Eq.~\eqref{eq:coll_decay}. This means it is not restricted to cavity systems but can in principle also be achieved with other systems featuring this dynamic, e.g.\ emitters coupled to a waveguide.} 

For a single quantum trajectory without a jump (no-click limit~\cite{garraway1994evolution,zerba2023measurement,tomasi2024stable,gal2024entanglement,DiFresco2024metrology}), the dynamics of the above described collective decay is governed by the non-Hermitian Hamiltonian 
\begin{align}
    \hat{H}_{\mathrm{nh}} = - i \frac{\gamma}{2} \hat S_+\hat S_-\, .
    \label{eq:nonhermitianH}
\end{align}
Note that the no-click trajectories are always the same  {but require post-selection, therefore we focus solely on them in the following (see Fig.~\ref{fig:setup_sketch}).} The eigenstates of this Hamiltonian are the symmetric Dicke states~\cite{dicke1954coherence}, denoted by $|m \rangle \equiv | S = N/2, S_z = m \rangle$,  with $m=-N/2,\ldots,N/2$, which are simultaneous eigenstates of the collective spin operators $\hat S_z$ and $\hat {\mathbf S}^2 = \hat S_x^2 + \hat S_y^2 + \hat S_z^2$.
This can be seen by rewriting the non-Hermitian Hamiltonian in the following way 
\begin{align}\label{eq:nonhermitianH2}
    \hat H_{\mathrm{nh}} = - i \frac{\gamma}{2} \hat S_+\hat S_- = - i \frac{\gamma}{2} (\hat {\mathbf S}^2 -\hat S_z^2 + \hat S_z),
\end{align}
where we have used $\hat S_\pm \equiv \hat S_x \pm i\hat S_y$. By acting on the Dicke states with Hamiltonian~\eqref{eq:nonhermitianH2}, we obtain
\begin{align}
    \hat H_{\mathrm{nh}}|m\rangle = - i \frac{\gamma}{2} \left(\frac{N (N + 2)}{4} - m^2 + m \right) |m\rangle  \equiv \varepsilon_{m} |m\rangle,
\end{align}
which describes how these states decay in time.
The Dicke states that decay at the slowest rate are those with the smallest (non-zero) modulus of the purely imaginary eigenvalues, $\Gamma_m = |\varepsilon_m|$. In our case, the minimal value, $\min_m \Gamma_{m} = \gamma N / 2$, is reached for $m = -N/2+1$ and $m = N/2$, which corresponds to the first excited state and the maximally excited spin state, respectively.  {This implies that there exists a set of initial states whose non-Hermitian dynamics passes through highly entangled states, provided no jumps occur for a sufficiently long time.} In particular, the maximally entangled cat state  {can be conditionally generated} during the non-Hermitian time evolution if, at a specific moment $t_\mathrm{c}$, the populations of the extreme spin states $m = \pm N/2$ are equal, while the population of all other spin states is negligible:
\begin{align}\label{eq_sm:condition_cat} | \langle -N/2 | \psi (t_\mathrm{c}) \rangle |^2 = | \langle N/2 | \psi (t_\mathrm{c}) \rangle |^2 \gg | \langle m | \psi (t_\mathrm{c}) \rangle |^2, \hspace{0.5cm} \end{align}
for all $m \neq \pm N/2$. Since the population in state $| m \rangle$ decreases as
\( | \langle m | \psi (t) \rangle |^2 = \exp(-\Gamma_m t ) | \langle m | \psi_0\rangle |^2,
\)
where $|\psi_0\rangle$ is the initial state, the equality in Eq.~\eqref{eq_sm:condition_cat} gives
\begin{align} e^{-\gamma N t_\mathrm{c} } | \langle N/2 | \psi_0 \rangle |^2 = | \langle -N/2 | \psi_0 \rangle |^2,
\end{align}
and, therefore,
\begin{align}\label{eq:cattime} t_\mathrm{c} = \frac{1}{\gamma N} \ln \left[ \frac{ | \langle N/2 | \psi_0 \rangle |^2 }{ | \langle -N/2 | \psi_0 \rangle |^2 } \right],
\end{align}
 {which defines the time at which a cat state can be generated if the condition in Eq.~\eqref{eq_sm:condition_cat} is satisfied. Note that, while the absence of photon emission in the no-click trajectory until $t_c$ might suggest that superradiance is irrelevant, it is precisely the non-Hermitian superradiant dynamics that enables the generation of highly entangled states.}

Several conclusions can be drawn from these considerations. First, as expected from superradiance, the time $t_\mathrm{c}$ strongly depends on the number of atoms.  { This collective effect enables the generation of highly entangled states on a timescale much shorter than any other relevant one, where large atom numbers are even beneficial}. Second, although the initial population of the excited state must exceed that of the ground state, the latter cannot be exactly zero, i.e., $| \langle -N/2 | \psi_0 \rangle |^2 \neq 0$. Finally, the time $t_\mathrm{c}$ cannot be too large, as no quantum jumps must occur during the evolution up to $t_\mathrm{c}$. Nevertheless, the probability of avoiding a jump during the evolution to $t_\mathrm{c}$ is generally very low.  {Thus, trajectories leading to the maximally entangled cat state are extremely rare. In the following, we demonstrate how to significantly enhance this probability by optimizing the initial state of the atomic ensemble.}


 {\emph{Initial state optimization with OAT---}}
In order to optimize the probability of generating highly entangled states, we use OAT 
to prepare the initial state. This can increase the probability for trajectories to reach a highly entangled state before a photon is emitted. This step can be understood as entanglement seeding,  {where the initial entanglement, set by OAT, is amplified. Postselection on no-click trajectories allows us to identify these states.} Although other methods can be used to seed the entanglement, such as measurement-induced squeezing~\cite{feedbacksqueezing2002wiseman,conditionalsqueezing2011thompson,Bohnet2014reduced,thompson2016mesuremetnsqueezing} or two-axis counter-twisting~\cite{twoaxistwisting2011Liu,kajtoch2015TACT,gediminas2022twoaxis}, OAT is particularly relevant due to its accessibility in state-of-the-art cavity-QED experiments~\cite{leroux2010cavitysqueeezing,pedrozo2020entanglement, colombo2022timereversal, vuletic2022spinssuqeezingOAT,barbarena2024tradeoffsqueezing,SM}.
The OAT unitary evolution reads
\begin{align}\label{eq:OAT_unitary}
    \hat U(\chi) = \exp\left( - i \chi \hat S_x^2\right).
\end{align}
We want to note here that OAT under idealized conditions is itself capable of generating the maximally entangled cat state~\cite{gietka2015twistedstates} for the twisting parameter $\chi=\pi/2$  {(and an even number of atoms~\cite{SM}). However, its experimental realization is very challenging, especially for macroscopic systems.}



\begin{figure}[tb!]
    \centering
    \includegraphics[width=0.48\textwidth]{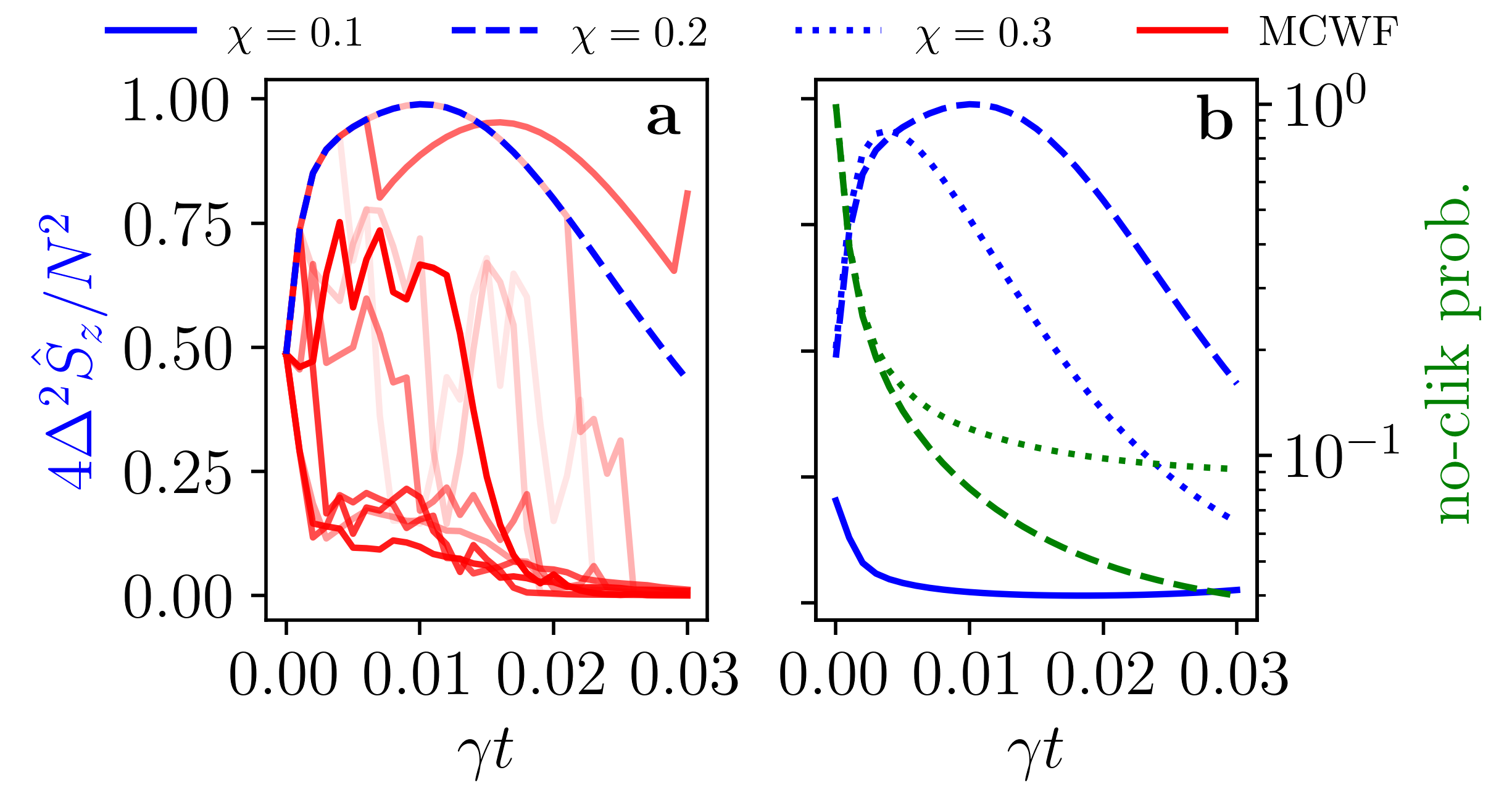}
    \caption{ {The non-Hermitian no-click trajectory (blue line) and exemplary Monte-Carlo wavefunction (MCWF) trajectories (red lines)} for the squeezed inverted ensemble of $N=100$ atoms. 
    In (a), we show quantum trajectories for a significant amount of entanglement in the initial state ($\chi = 0.2$). Out of ten randomly generated trajectories (shaded red lines), one reaches the highest entangled state allowed by the no-click trajectory (dashed blue line). The trajectories depict the variance of $\hat S_z$ which is a measure of entanglement between the atoms for pure states. For $N=100$ and the considered non-Hermitian dynamics, $\Delta^2 \hat S_z \equiv \langle \hat S_z^2\rangle - \langle \hat S_z\rangle^2 = N^2/4=2500$ indicates the maximally entangled state.
    In (b), we show the $\hat S_z$ variance versus time of no-click trajectories for three values of $\chi$. If the initial entanglement is too low ($\chi = 0.1$), the probability of generating a highly entangled state is negligible because the process takes too long. 
     {The no-click probability (green lines) decreases fast with time. Amplifying the entanglement less by stopping the time evolution earlier can strongly increase the success (no-click) probability.}} 
    \label{fig:trajectories}
\end{figure}

 {
To quantify the entanglement we employ the variance of the $\hat S_z$ operator. Although other measures could be used to quantify entanglement \cite{SM}, our choice of variance is motivated by experimental accessibility. This quantity is particularly useful in the context of quantum metrology since a large variance value directly corresponds to a large quantum Fisher information~\cite{smerzi2018rmpmetorlogy}.} 

 {Fig.~\ref{fig:trajectories}(a) shows ten example Monte-Carlo wavefunction trajectories for the twisting parameter $\chi=0.2$. This initial state is represented by the first Bloch sphere in Fig.~\ref{fig:setup_sketch}(b). Most trajectories exhibit at least one photon emission, which means their state will be discarded. However, one of the trajectories features no quantum jump until the peak value of the variance is reached, i.e.\ it overlaps with the curve of the non-Hermitian dynamic (dashed blue line). Note that in order to stop the non-Hermitian evolution at the desired instant of time, the atoms need to be decoupled from the superradiant dynamics. In the above-described cavity setup, this can e.g.\ be achieved by abruptly detuning the cavity from the atoms or transferring the excited atomic states to another long-lived (decoupled) state with a fast $\pi$-pulse.}

 {Fig.~\ref{fig:trajectories}(b) shows the dynamics according to the non-Hermitian Hamiltonian~\eqref{eq:nonhermitianH} for $N=100$ excited atoms with different initial squeezing according to Eq.~\eqref{eq:OAT_unitary}. We see that for specific twisting parameters ($\chi = 0.2$ and $0.3$) the variance is strongly increased, for $\chi = 0.2$ it even approximately reaches the maximal variance. 
For $\chi=0.1$ the variance decreases at the beginning, which corresponds to the collective state becoming almost the fully excited state. In this scenario, the initially seeded entanglement was too little to be amplified. 
The figure also shows that there is a specific time when the variance reaches its peak in the non-Hermitian dynamics. In the following, we will refer to this time as the optimal time $t_\mathrm{opt}$, see also Fig.~\ref{fig:setup_sketch}(c). It is also worth mentioning that if one does not necessarily have to reach the peak, it can be more efficient to stop the time evolution with the non-Hermitian Hamiltonian earlier, which makes a no-click trajectory more probable, see Fig.~\ref{fig:trajectories}(b).}



\begin{figure}[tb]
    \centering
    \includegraphics[width=0.49\textwidth]{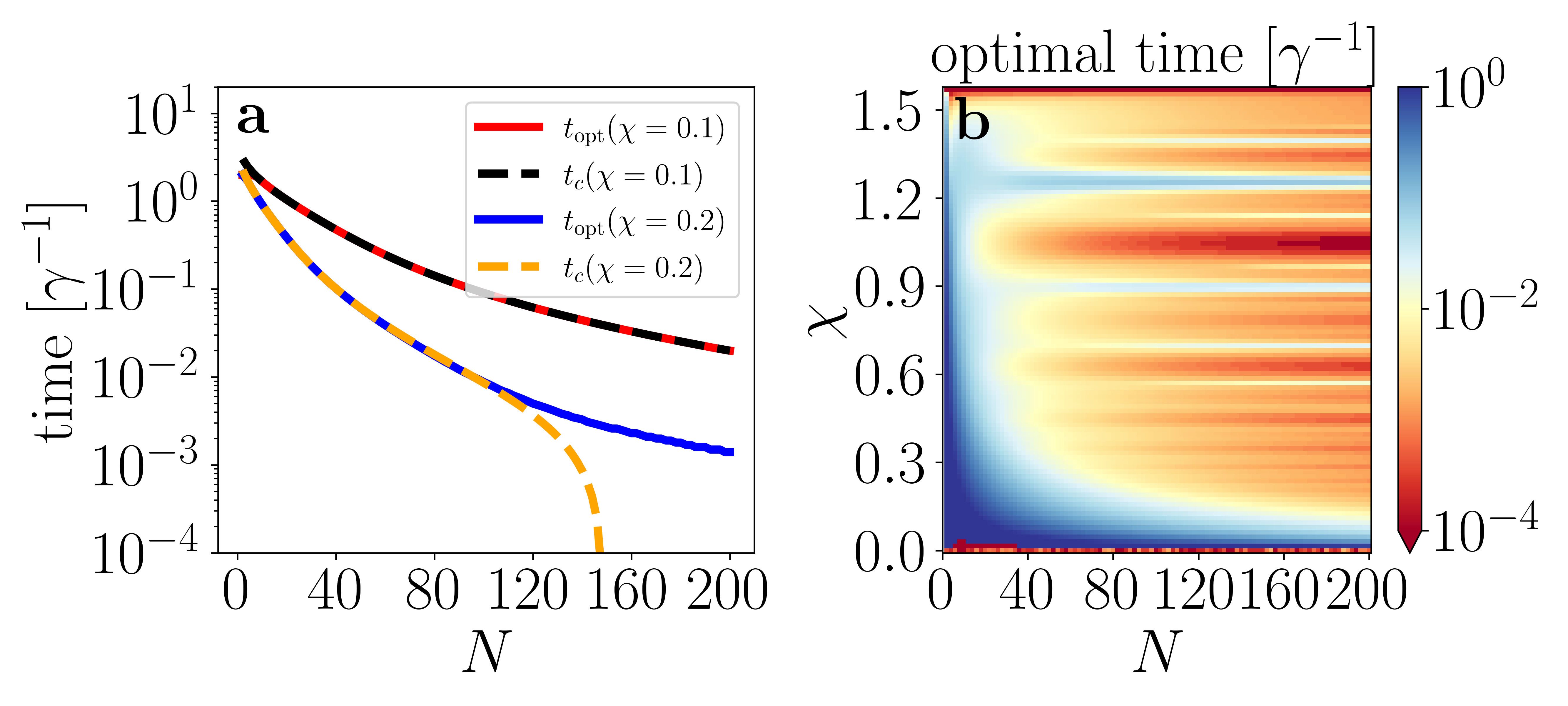}
    \caption{Comparison between $t_\mathrm{c}$  {[Eq.~\eqref{eq:cattime}]} and the time at which maximal entanglement is  {actually} generated $t_\mathrm{opt}$ for a no-click trajectory. (a)  {Numerically calculated} optimal time (solid lines) and $t_\mathrm{c}$ (dashed lines). 
    If the squeezing is too strong (visible for  {$N > 100$ and} $\chi = 0.2$), the initial state does not satisfy the requirements for the cat state generation, as at $t=0$ the probabilities of the system to be in the ground state and the excited state are almost equal. 
    (b) Optimal time for the maximal entanglement generation (not necessarily a cat state) as a function of $N$ and $\chi$.}
    \label{fig:Atimes}
\end{figure}

 {
In the following, we investigate the peak variance of $\hat S_z$ and the corresponding no-click probability depending on the twisting parameter $\chi$ and the atom number $N$. 
Since the probability of a photon emission increases with time, the above described optimal time $t_\mathrm{opt}$, i.e.\ when the peak variance of $\hat S_z$ is reached, is a crucial quantity of the dynamics. 
We expect the optimal time to align with the analytically derived time $t_{c}$ at which the maximally entangled cat state can be generated [see Eq.~\eqref{eq:cattime}], at least for parameters where a cat state is approximately reached. In Fig.~\ref{fig:Atimes}(a) we compare $t_\mathrm{opt}$ and $t_{c}$. 
For weak squeezing ($\chi = 0.1$), the optimal time aligns with the prediction from Eq.~\eqref{eq:cattime}. However, as the squeezing increases ($\chi = 0.2$) and the number of atoms $N$ grows, we enter a strong over-squeezing regime~\cite{ueda1993SqueezedSpinStates,oversqueezed2014oberthaler, oversqueezed}, where the initial state already has an almost equal occupation of the $m = \pm N/2$ states. In this regime, $t_c$ approaches zero. 
Fig.~\ref{fig:Atimes}(b) shows the full scan over $N$ and $\chi$ for the numerically calculated optimal time. We see that there is a threshold twisting parameter for the process to be fast which reduces with $N$. Note that a twisting parameter of $\chi = \pi/2$ corresponds to an initial cat state. 
} 


\begin{figure}[tb]
    \centering
    \includegraphics[width=0.5\textwidth]{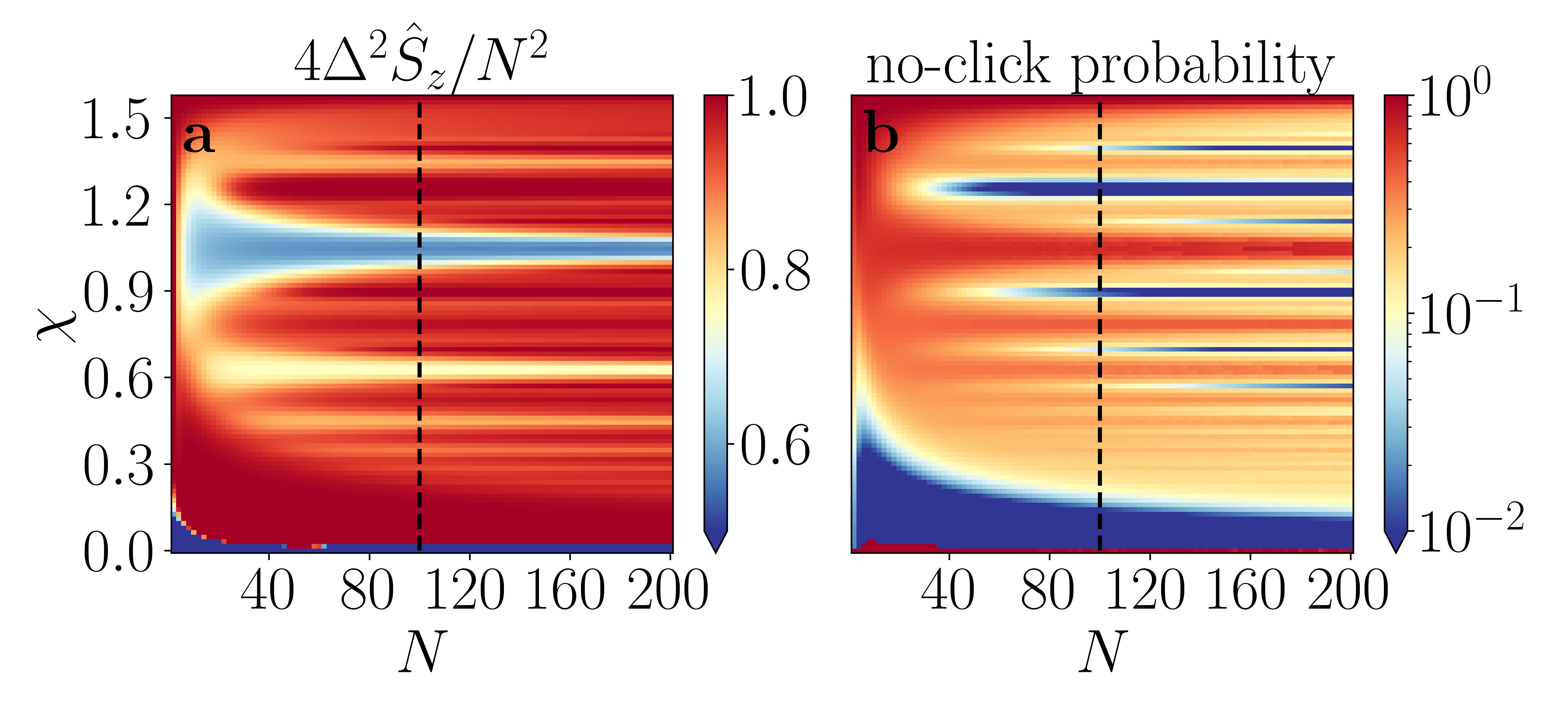}
    \includegraphics[width=0.44\textwidth]{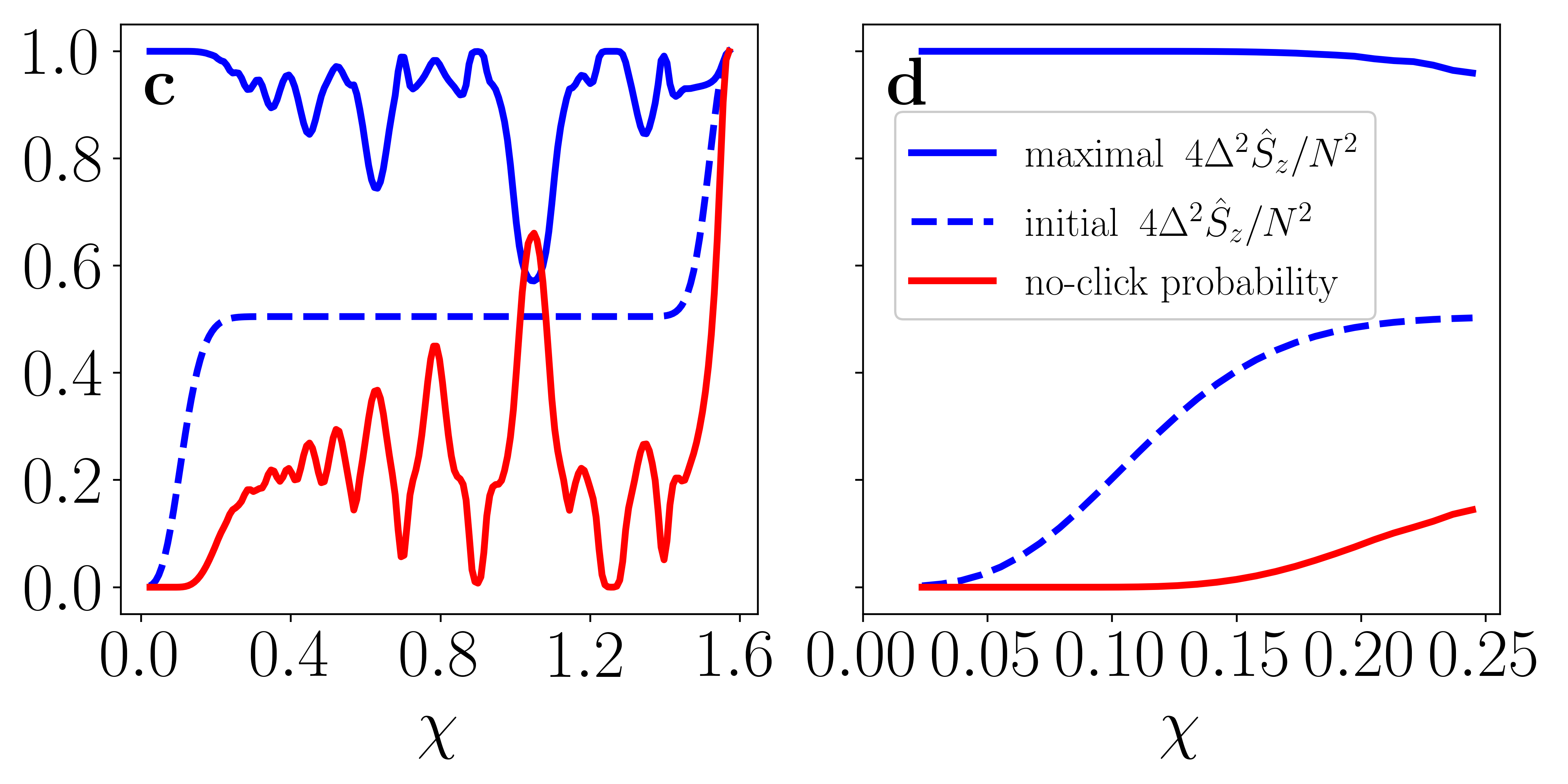}
    \caption{Optimizing the probability of generating highly entangled states. In (a), we show the normalized maximal variance of $\hat S_z$ for no-click trajectories {(i.e.\ at $t_\mathrm{opt}$)} as a function of $\chi$ and (even) number of atoms $N$. (b) Probability of obtaining the state with maximal variance from (a). Panels (c) and (d) represent a cut through $N=100$ for the maximal variance of $\hat S_z$ (solid blue) and the no-click probability (solid red). The dashed blue line shows the initial variance after the OAT squeezing.  { In (d), we zoom at the optimal operating point, where the twisting parameter $\chi$ is not too large.}}
    \label{fig:scan1}
\end{figure}


The no-click probability at time $t$ in the non-Hermitian dynamics is given by the state's absolute square of the norm $| \langle \psi_\mathrm{nh}(t) | \psi_\mathrm{nh}(t) \rangle |^2$ with $|\psi_\mathrm{nh}(t)\rangle = \exp(-i t \hat H_\mathrm{nh})|\psi_0\rangle$~\cite{dalibard1992wavefunction,dum1992monte,molmer1993monte}. 
In Fig.~\ref{fig:scan1} we present the results of the numerical simulations for the non-click probability and the peak value of the $\hat S_z$ operator variance to quantify the entanglement. 
Specifically, in Fig.~\ref{fig:scan1}(a), we plot the peak value of the $\hat S_z$ operator variance normalized to the maximal value ($N^2/4$) as a function of the initial twisting strength $\chi$ for various numbers of atoms $N$ {, see~\cite{SM} for a discussion on the atom number parity dependence.} Note that a cat state is obtained once the normalized variance is equal to~1. As can be seen, most no-click trajectories where the entanglement has been sufficiently seeded lead to highly entangled states. In Fig.~\ref{fig:scan1}(b), we plot the probability of observing a trajectory where no jump occurs until the  {peak of the $\hat S_z$ operator variance is reached, which is given by the norm of the state at the above-described optimal time $t_\mathrm{opt}$.} 
Importantly, even for slightly over-squeezed states, the probability of generating even more  {entanglement} can be significant.  {We also want to emphasize the qualitative agreement between the optimal time in Fig.~\ref{fig:Atimes}(b) and the no-click probability in Fig.~\ref{fig:scan1}(b). This highlights the intuition that a high success probability can only be achieved if the time to reach the state is short.} 
In Fig.~\ref{fig:scan1}(c) we show a cut through $N=100$ for the scans in  {Fig.~\ref{fig:scan1}(a) and (b), while in Fig.~\ref{fig:scan1}(d) we zoom in on the most promising parameter regime.} The plotted initial variance (blue dashed line) indicates that stronger squeezing with OAT does not increase the variance of $\hat S_z$ above a certain value ($\chi \approx 0.2$ for $N=100$ atoms). Only at a twisting parameter $\chi$ close to $\pi/2$ it grows again rapidly as the initial state is itself very close to the cat state. In general, for the best performance of the protocol the initial state needs to be slightly over-squeezed as can be seen in Fig.~\ref{fig:scan1}(c) and (d). For $100$ atoms, a good operating point is at a twisting parameter of $\chi \approx 0.2$, which yields a no-click probability of $7.3\%$. The corresponding cat state fidelity is $90\%$. 
 {We want to emphasize here that Fig.~\ref{fig:Atimes} and Fig.~\ref{fig:scan1} also indicate a beneficial scaling with the atom number. While the variance stays approximately constant, the optimal time reduces and the probability increases for a specific $\chi$ (in the promising parameter regime).}

 {\emph{Conclusions---}}In summary, we have presented a superradiance-based method that allows  {for the conditional generation of} highly entangled states, including the maximally entangled cat state. The method relies on harnessing dissipation and maximizing the probabilities of quantum trajectories leading to highly entangled states.  {To achieve this, we analyzed} an effective non-Hermitian Hamiltonian which governs the dynamics between quantum jumps. By finding its eigenvalues in the Dicke basis, we  {identified} which initial states increase the probability of generating highly and maximally entangled states in the no-click limit. Subsequently, by exploiting experimentally accessible OAT, we have shown how one can generate these states, including cat states, using initially squeezed and over-squeezed states. 

Our work constitutes an important step toward generating macroscopic  {highly} entangled states of atoms  {on time scales much faster than spontaneous emission, a significant improvement over existing techniques that typically require longer evolution times or are hindered by decoherence.} More generally, this framework can be applied to any collective (pseudo) spin coupled to a strongly damped harmonic oscillator,  {offering a versatile approach compared to methods such as one-axis twisting under idealized conditions, which often struggle with scalability and decoherence in large systems. Unlike purely dissipative state preparation methods, which rely on steady-state engineering and are limited by the trade-off between entanglement and dissipation, our protocol exploits transient dynamics to achieve highly entangled states efficiently.} Furthermore, although our idea is based on high-efficiency photon detectors, we stress that { 
neither them nor experimental conditions} have to be ideal \cite{SM},  {making the approach experimentally accessible with current technologies.}


\acknowledgements 
 {\emph{Acknowledgements---}}The authors would like to thank Tommaso Pirozzi and Bruno Vinciguerra for fruitful discussions, Marcin P\l{}odzie\'n for carefully reading the manuscript and the Aurora Excellence Fellowship Program for support. C.H. was supported by the Carlsberg Foundation through the “Semper Ardens” Research Project QCooL. Simulations were performed using the open-source \textsc{QuantumOptics.jl}~\cite{kramer2018quantumoptics} framework in \textsc{Julia}. This research was funded in part by the Austrian Science Fund (FWF) [grant DOIs: 10.55776/ESP171 and 10.55776/M3304]. For open access purposes, the authors have applied a CC BY public copyright license to any author accepted manuscript version arising from this submission. The data presented in this article is available from \cite{zenodo}.



\begin{thebibliography}{90}
\bibitem [{ove()}]{oversqueezed}%
  \BibitemOpen
  \href@noop {} {}\bibinfo {note} {A state is called over-squeezed when it starts to wrap around the Bloch sphere and hence can no longer be described by a Gaussian state. In the context of OAT, a state is over-squeezed if the twisting parameter $\chi$ is greater than the optimal twisting parameter $\chi_0 = 6^{1/6} N^{-2/3}$.}\BibitemShut {Stop}%
\end{thebibliography}

%

\newpage
\onecolumngrid
\section{Supplemental Material: \\ Conditional {Entanglement Amplification} via {Non-Hermitian} Superradiant {Dynamics}}
 



\setcounter{equation}{0}
\setcounter{figure}{0}
\renewcommand{\theequation}{S\arabic{equation}}
\renewcommand{\thefigure}{S\arabic{figure}}




\section{Atom number parity}
\begin{figure}[htb!]
    \centering
    \includegraphics[width=0.5\textwidth]{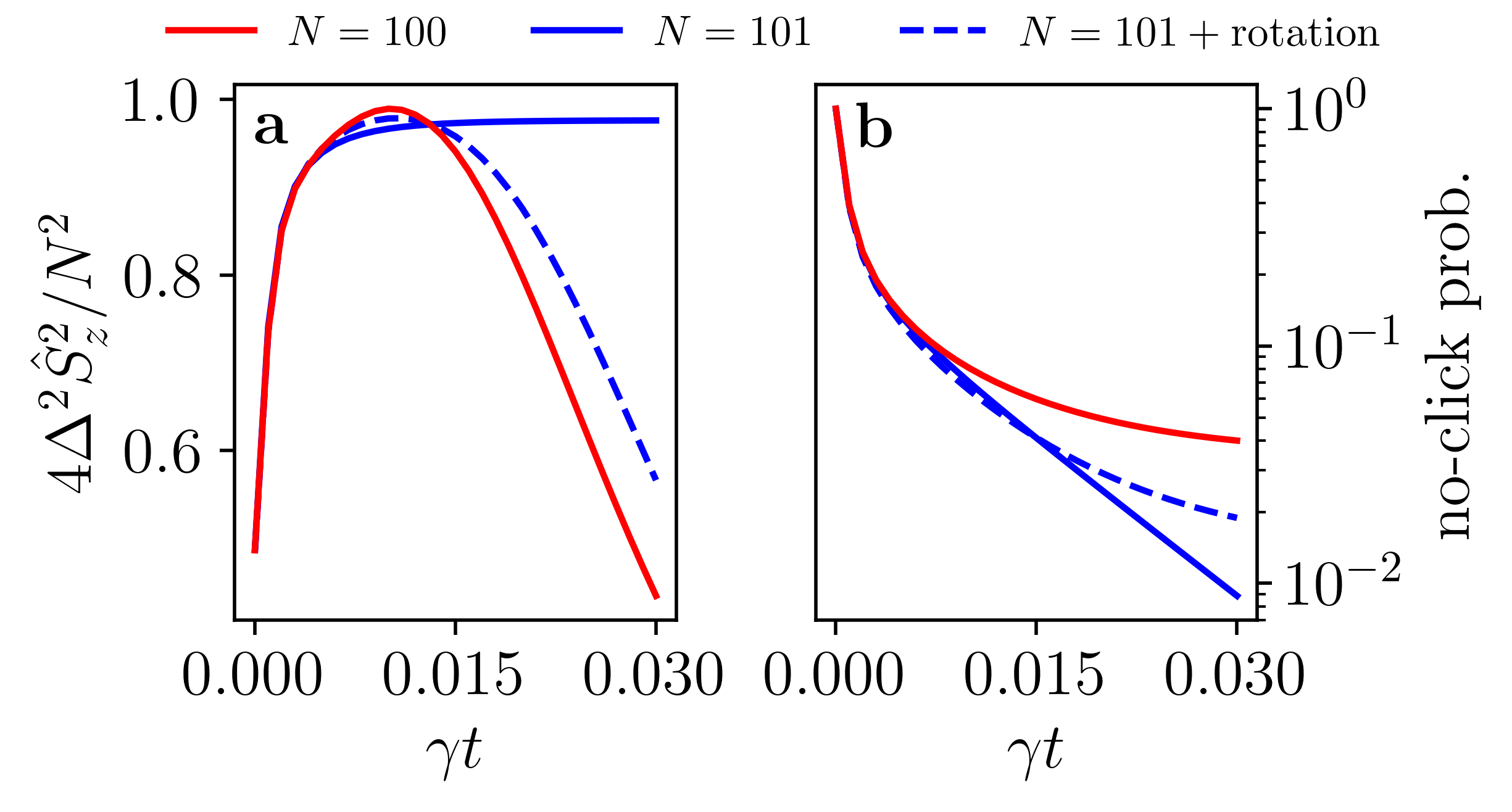}
    \caption{{Atom number parity dependence. (a) Time evolution of the normalized variance for $N=100$ (red) and $N=101$ (blue). The two cases differ mainly after the optimal time. A small rotation around the $y$-axis ($\theta = 0.1$) allows to more closely resemble the time evolution of the even atom number (dashed blue line). (b) No-click probabilities corresponding to trajectories in (a).  The change for $N=100$ with the small rotation is barely visible, therefore we do not show it. The squeezing parameter is $\chi=0.2$ for all plots.}}
    \label{fig:timeevolution_odd}
\end{figure}

{In the main text, we only considered even numbers of particles. The entanglement amplification, however, can depend on the parity of the atom number. For example, in order to reach the maximally entangled cat state, the parity of the atom number is crucial even for the one-axis Twisting Hamiltonian as it conserves the parity. Therefore an odd number of particles does not lead to the occupation of the ground state. Specifically, the lowest Dicke state that can be occupied by an OAT dynamics of a fully inverted ensemble with an odd number of atoms is the first excited Dicke state $| -N/2+1 \rangle$. 
In Fig.~\ref{fig:timeevolution_odd} we show the time evolution of the variance and the no-click probability for $N=100$ and $N=101$ with a twisting parameter of $\chi = 0.2$. Up to the optimal time (for $N=100$), we see only a slight difference in the variance and the no-click probability. After the optimal time, however, the variance does not reduce for the odd number of atoms (blue solid line), also the no-click probability continues to decrease faster. This is due to the fact that the ground state can not be populated with OAT for an odd number of atoms if the initial state is the fully inverted ensemble. Note that this also means that the optimal time is not properly defined for an odd number of atoms since there is no maximum for the variance. 
With an extra rotation around the $y$-axis on the generalized Bloch sphere generated by $\exp(- i \theta \hat S_y)$ and a subsequent evolution by OAT (or vice versa), the ground state can also be populated. This allows to closely resemble the even case with an odd number of atoms (dashed blue line, $\theta=0.1$). This can be useful if the atom number is uncertain. Note that even without the rotation the no-click trajectory leads to highly entangled states. Furthermore, in order to create a cat state the atom number anyway needs to be known with certainty. 
}

{In principle, the use of alternative entanglement seedings mechanisms could reduce the sensitivity to the even-odd parity of the atom number. However, from an experimental standpoint, many-body entanglement is most commonly realized through spin-squeezing. Therefore, a practical approach to mitigating the parity dependence would likely involve optimized squeezing protocols combined with suitable rotations. }

\section{Cat state fidelity and parameter uncertainty}
\begin{figure*}[h!]
\centering
\includegraphics[width=0.9\textwidth]{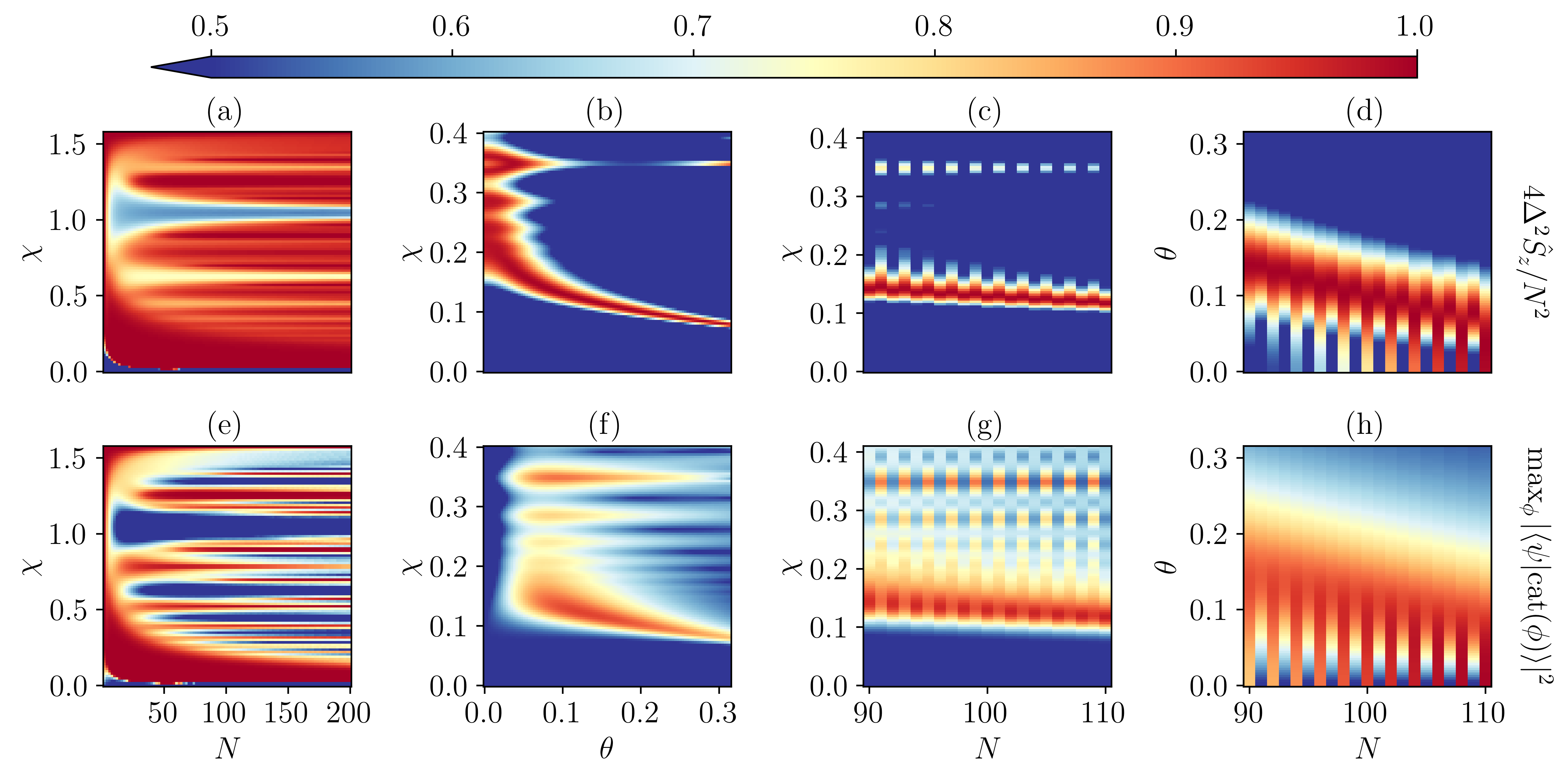}
\caption{{Comparison between the amount of entanglement as measured by the variance of $\hat S_z$ and cat state fidelity. (a)-(d) depict the normalized variance of $\hat S_z$. (e)-(h) show the cat state fidelity. The parameters when kept constant are $\theta = 0$ [(a) and (e)], $N = 101$ [(b) and (f)], $\theta = 0.1$ [(c) and (g)] and $\chi = 0.2$ [(d) and (h)]. Note that a state can be highly entangled but has an almost vanishing overlap (fidelity) with a cat state. For the optimal cat state generation, the initial state should be strongly squeezed or weakly over-squeezed.} }
\label{fig:Afid}
\end{figure*}

In the main text, we used the variance of the $\hat S_z$ operator to quantify the amount of entanglement in the optimally entangled state generated through no-click trajectories. Here, we present additional simulations that show the fidelity of generating a maximally entangled state. To this end, we calculate the overlap of a cat state with phase $\phi$
\begin{align}
    \mathcal{F} = \max_{\phi} |\langle \psi|\mathrm{cat}(\phi)\rangle|^2
\end{align}
and plot the maximal value of the overlap. We need to maximize over the phase $\phi$ because the cat state phase depends on the initial conditions such as the amount of squeezing, optional rotation angle, and number of atoms. The results are presented in Fig.~\ref{fig:Afid}. 
{In (e) we see that a high cat state fidelity can be achieved. In this case we use the optimal time $t_\mathrm{opt}$ for each data point.}

{
Furthermore, we also show the influence of uncertain parameters, i.e.\ $N$, $\chi$ and $\theta$. In Fig.~\ref{fig:Afid}(b)-(d) and (f)-(h), we do not optimize the time but choose one final time $t_\mathrm{end} = 0.033 / \gamma$ of the non-Hermitian superradiant dynamics in all scans, corresponding to $t_\mathrm{opt}$ for $\chi = 0.2$, $N=100$ and $\theta = 0$.  
The cat state fidelity has a stronger dependence on $N$ compared to the variance of $S_z$. However, this is not surprising since the atom number needs to be precisely known to produce a cat state, regardless of the method. 
We also see that a slight rotation of the collective ensemble on the Bloch sphere can remove differences in trajectories for over-squeezed states. 
}

\begin{figure*}[h!]
\centering
\includegraphics[width=0.6\textwidth]{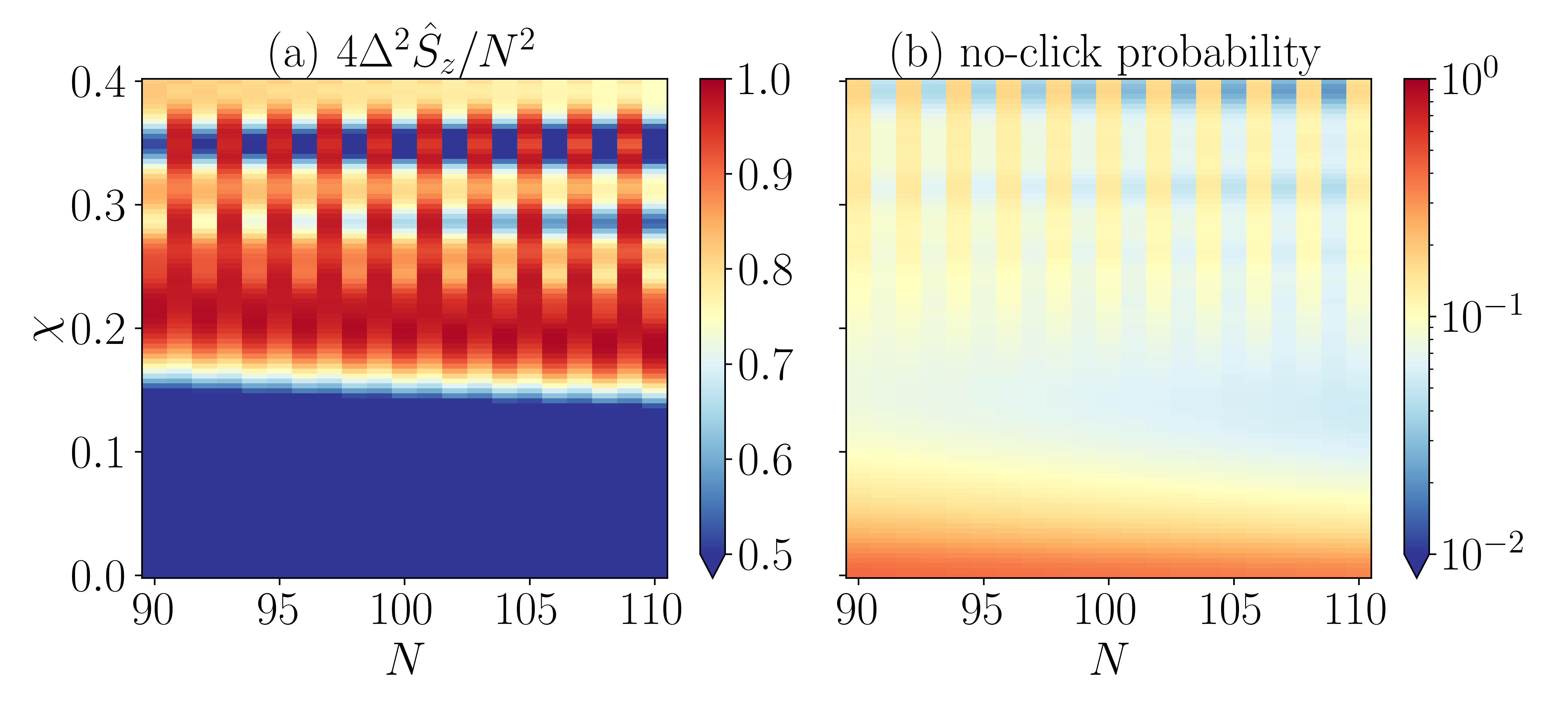}
\caption{{Parameter uncertainty. (a) depicts the variance and (b) the no-click probability for uncertainties in $N$ and $\chi$. One final time $t_\mathrm{end} = 0.033/\gamma$  of the superradiant dynamics is used for all parameters in the plot and there is no rotation $\theta = 0$.} }
\label{fig:param_unc}
\end{figure*}

In Fig.~\ref{fig:param_unc} we show the variance (a) and the no-click probability (b) for uncertain parameters for $\theta = 0$ and the same final time $t_\mathrm{end} = 0.033/\gamma$  for all parameters, which is the optimal time for $N = 100$, $\chi = 0.2$ and $\theta = 0$. We see that uncertainty in $N$ and $\chi$ (around $N=100$ and $\chi = 0.2$) does influence the variance and the no-click probability, but the protocol can still significantly enhance the entanglement. This shows that an imperfect OAT protocol can still be used to seed the entanglement for the amplification. 

\section{Experimental realization} 
In this section, we briefly discuss potential experimental implementations. Ideal platforms to generate macroscopic cat states are state-of-the-art cavity-QED experiments, where squeezing and superradiance are possible. The most challenging part with such setups is the different parameter regime for the OAT squeezing protocol and superradiance. Cavity-mediated squeezing usually requires a strong collective coupling with resolved normal-mode splitting peaks ($\sqrt{N}g > \kappa$)~\cite{schleier2010squeezing, leroux2010cavitysqueeezing, conditionalsqueezing2011thompson, Bohnet2014reduced, braverman2019unitary, li2022collective}. 
In contrast, for superradiance the bad cavity regime ($\sqrt{N}g \ll \kappa$) is needed~\cite{dicke1954coherence,norcia2016superradiance,norcia2018frequency,laske2019pulse,hotter2023cavity,bohr2024collectively}. One possibility to attain both regimes with one cavity setup is to use a strong atomic transition for squeezing and then transfer it to a narrow atomic (clock) transition~\cite{pedrozo2020entanglement} for superradiance. Strontium-87 or Ytterbium-171 e.g.\ feature such transitions. For optical transitions, however, the different wavelengths make it non-trivial to achieve equal coupling for both transitions in a standing wave cavity. For this reason, it would be beneficial to use a ring cavity where the atom-cavity coupling strength can be equal for all atoms. Another option would be to use a separate cavity for superradiance or to work in the microwave regime where these issues are not relevant due to the much longer wavelengths.

\section{Detector efficiency}
\begin{figure*}[htb!]
    \centering
    \includegraphics[width=0.4\textwidth]{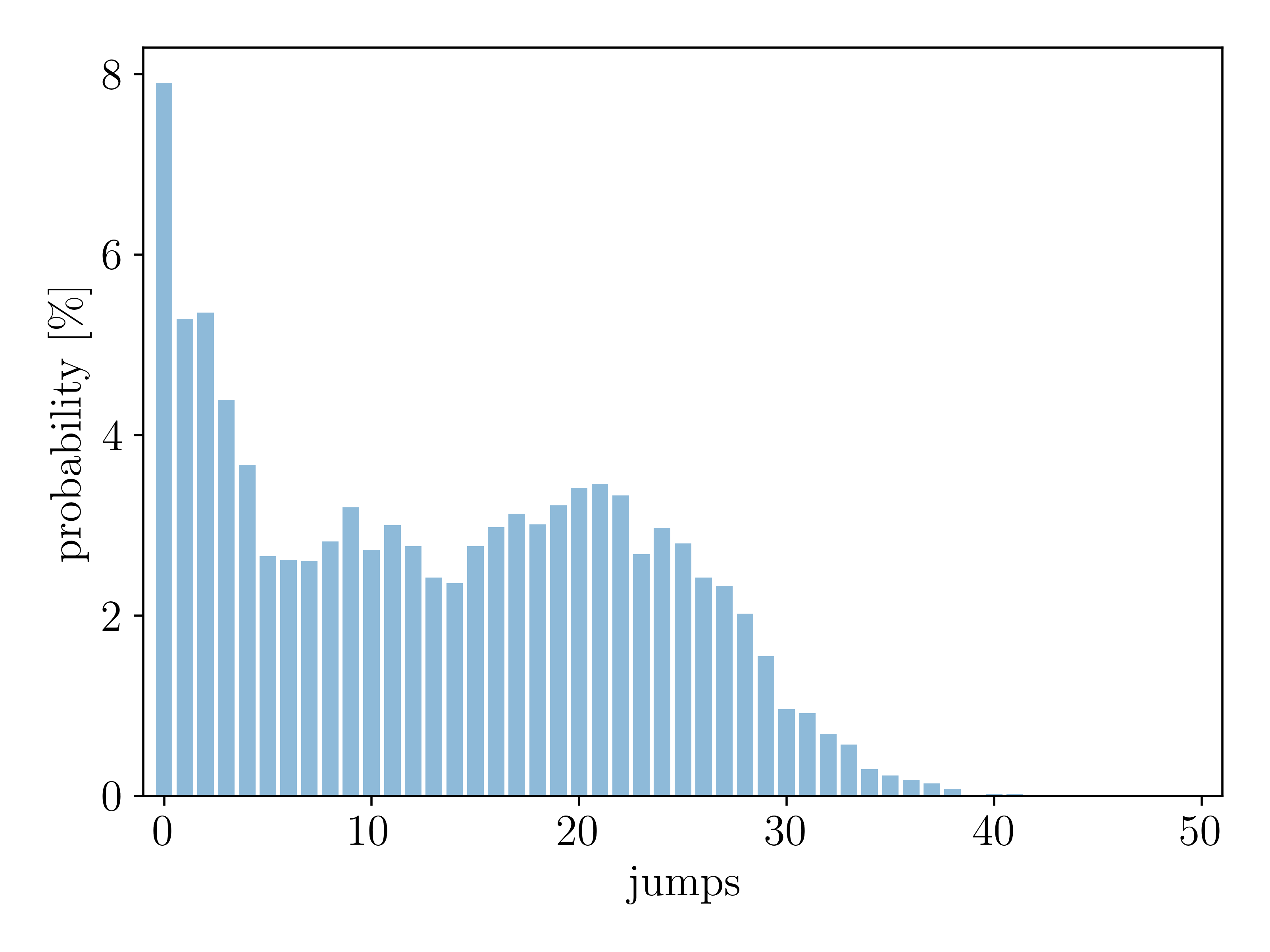}
    \caption{Jump statistic. The histogram shows the probability of $n$ quantum jumps to occur until a cat state would be generated in a no-jump trajectory. For a detector with an efficiency of $\eta=90\%$ the corresponding precision (no jump in a no-click trajectory) is $~93\%$. The parameters are $N = 100$ and $\chi=0.2$.}
    \label{fig:jump_stats}
\end{figure*}

We briefly elaborate on the influence of a non-perfect photon detector. Fig.~\ref{fig:jump_stats} shows the probability $p_n$ of $n$ quantum jumps occurring until the time a cat state would be generated for a no-jump trajectory. In a simplified model a photon detector with an efficiency $\eta$ has a probability of $(1-\eta)^n$ to miss $n$ photons. This means the precision $p_0 / \sum_{n=0}^N p_n (1 - \eta)^n$ describes the probability of actually having no jump in a no-click trajectory (true positive / all positives). For a high-fidelity photon detector with an efficiency of $\eta = 90 \%$ we obtain a precision of $\sim 93 \%$ for the distribution in Fig.~\ref{fig:jump_stats}. The jump probabilities are simulated with $10^4$ MCWF trajectories. 

\end{document}